\documentclass[twocolumn]{aastex631}
\usepackage{amsmath,amssymb}

\makeatletter
\renewcommand*{\env@matrix}[1][*\c@MaxMatrixCols c]{%
  \hskip -\arraycolsep
  \let\@ifnextchar\new@ifnextchar
  \array{#1}}
\makeatother

\usepackage{nicematrix}
\usepackage{xspace}
\usepackage{units}
\usepackage{subfigure}

\newcommand{\lldf}{\ensuremath{L_{\text{LDF}}}}
\newcommand{\xmax}{\ensuremath{X_{\text{max}}}}

\newcommand{\fluence}{\ensuremath{\mathcal{F}}\xspace}
\def\diff{\mathrm{d}}
\makeatletter
\newcommand{\ufrac}[2]
{
 \frac{#1}{\mathpalette\ufrac@den{#2}}%
}
\newcommand{\ufrac@den}[2]{#1#2}
\newcommand{\photonsignificance}{\ensuremath{\bar{P}_\text{bg}}}
\makeatother

\submitjournal{ApJ}

\begin{document}

\title{Search for UHE Photons from Gravitational Wave Sources with the Pierre Auger Observatory}

\author{A.~Abdul Halim}
\affiliation{University of Adelaide, Adelaide, S.A., Australia}

\author{P.~Abreu}
\affiliation{Laborat\'orio de Instrumenta\c{c}\~ao e F\'\i{}sica Experimental de Part\'\i{}culas -- LIP and Instituto Superior T\'ecnico -- IST, Universidade de Lisboa -- UL, Lisboa, Portugal}

\author{M.~Aglietta}
\affiliation{Osservatorio Astrofisico di Torino (INAF), Torino, Italy}
\affiliation{INFN, Sezione di Torino, Torino, Italy}

\author{I.~Allekotte}
\affiliation{Centro At\'omico Bariloche and Instituto Balseiro (CNEA-UNCuyo-CONICET), San Carlos de Bariloche, Argentina}

\author{K.~Almeida Cheminant}
\affiliation{Institute of Nuclear Physics PAN, Krakow, Poland}

\author{A.~Almela}
\affiliation{Instituto de Tecnolog\'\i{}as en Detecci\'on y Astropart\'\i{}culas (CNEA, CONICET, UNSAM), Buenos Aires, Argentina}
\affiliation{Universidad Tecnol\'ogica Nacional -- Facultad Regional Buenos Aires, Buenos Aires, Argentina}

\author{J.~Alvarez-Mu\~niz}
\affiliation{Instituto Galego de F\'\i{}sica de Altas Enerx\'\i{}as (IGFAE), Universidade de Santiago de Compostela, Santiago de Compostela, Spain}

\author{J.~Ammerman Yebra}
\affiliation{Instituto Galego de F\'\i{}sica de Altas Enerx\'\i{}as (IGFAE), Universidade de Santiago de Compostela, Santiago de Compostela, Spain}

\author{G.A.~Anastasi}
\affiliation{Osservatorio Astrofisico di Torino (INAF), Torino, Italy}
\affiliation{INFN, Sezione di Torino, Torino, Italy}

\author{L.~Anchordoqui}
\affiliation{Department of Physics and Astronomy, Lehman College, City University of New York, Bronx, NY, USA}

% ...

\collaboration{1000}{The Pierre Auger Collaboration}
\email{spokespersons@auger.org}

\setlength{\marginparwidth}{10em}

\begin{abstract}
A search for time-directional coincidences of ultra-high-energy (UHE) photons above \unit[10]{EeV} with
gravitational wave (GW) events from the LIGO/Virgo runs O1 to O3 is conducted with the Pierre Auger
Observatory.
Due to the distinctive properties of photon interactions and to the background expected from hadronic 
showers, a subset of the most interesting GW events is
selected based on their localization quality and distance.
Time periods of 1000~s around and 1 day after the GW events are analyzed.
No coincidences are observed.
Upper limits on the UHE photon fluence from a GW event are derived that are typically at 
$\sim \unit[7]{MeV\,cm^{-2}}$ (time period 1000~s) and $\sim \unit[35]{MeV\,cm^{-2}}$ (time 
period 1 day). Due to the proximity of the binary neutron star merger GW170817, the energy of the 
source transferred into UHE photons above \unit[40]{EeV} is constrained to be less than 20\% of its
total gravitational wave energy. These are the first limits on UHE photons from GW sources.
\end{abstract}

\section{Introduction}

With the first gravitational waves (GW) measured by the Advanced LIGO and Virgo detectors in
2015~\citep{abbott16a}, a new window to the universe has been opened. In addition, a new type of
transient astronomical object has been observed for the first time: the merging process of two compact
stellar mass objects (compact binary merger, CBM). Since the first measurement in 2015, three
observation runs (O1, O2 and O3) have been conducted with a total yield of 91 confident GW
observations. The sources of these signals turned out to belong to different groups including the
merging events of binary black holes (BBH), binary neutron stars (BNS), and neutron star -- black hole
(NSBH) systems~\citep{gwtc1,gwtc2,gwtc2.1,gwtc3}.

An extensive follow-up campaign in the electromagnetic domain revealed a coincident
kilonova event from the BNS merger GW170817 whereas no astrophysical neutrino signal has been
identified~\citep{abbott2017a}. This observation became a milestone of multimessenger astronomy
and the first multimessenger observation involving GWs. The acceleration mechanisms of cosmic rays
for such an event are being debated in the theoretical 
community~\citep{fang17a,kimura2018a,rodrigues2019a,decoene2020a}. Although no further
observations of electromagnetic or neutrino counterparts from other GW sources have been confirmed
so far, BBH and NSBH mergers are also being discussed as possible candidates for the acceleration of
ultra-high-energy (UHE) cosmic rays and, hence, potential sources of high-energy neutrinos and
photons~\citep{kotera2016a,murase2016a,mckernan2019a}.

With its design sensitivity at the highest energies in the cosmic ray spectrum above $\unit[10^{18}]{eV}$, 
the Pierre Auger Observatory~\citep{aab2015a} plays an important role in the
multimessenger follow-up campaign of GW sources~\citep{kampert2019a}. Constraints on the
production of UHE neutrinos by the source of GW170817 and the first BBH mergers detected during O1
have been obtained~\citep{aab2017c,albert2017a}, and a stacking analysis has been performed using
83 confident BBH merger observations aiming to constrain the neutrino emission from the source class
as a whole~\citep{abreu23a}. A first analysis of GW sources with respect to an UHE photon signal
using the data of the Pierre Auger Observatory is reported here. Although the attenuation length of UHE
photons is of the order of $\unit[10]{Mpc}$ due to interactions with the cosmic background radiation
fields~\citep{risse07a} -- mainly the cosmic microwave background (CMB) and the universal radio
background (URB) -- it turns out that the exposure of the Pierre Auger Observatory towards UHE
photons is large enough to potentially observe photons from sufficiently close sources. More distant
sources on the other hand can be used to probe the attenuation of UHE photons in the background
radiation fields and an observation of an UHE photon from such a source could point to new physics 
scenarios~\citep{fairbairn2011a,galaverni08a}. Focusing on the most promising sources while keeping
an open window for unexpected discovery, a selected set of GW sources will be analyzed here to reduce
the overall background from hadronic cosmic rays.

The paper is structured as follows. In Sec.~\ref{sec:photon_at_auger}, a summary of the method used
to search for UHE photons with the Pierre Auger Observatory is provided. In Sec.~\ref{sec:gw_data}, an
overview of the already concluded GW observation runs and the GW data relevant for this work is given.
A description of the GW selection strategy that is used to pick only the most relevant GW sources is
detailed in Sec.~\ref{sec:gw_event_selection} followed by a brief discussion in
Sec.~\ref{sec:sensitivity} of the signal sensitivity that can be achieved using that selection. The final
results of the analysis are presented in Sec.~\ref{sec:results} in the form of upper limits on UHE
photons from this selection of sources. Sec.~\ref{sec:conclusion} concludes with a short summary and
a comparison of our results to other search results from the literature.

\section{Ultra-high-energy Photon Search at the Auger Observatory}\label{sec:photon_at_auger}

The search for an UHE photon signal in coincidence with a GW is carried using data collected by the 
surface detector
array (SD) of the Pierre Auger Observatory~\citep{aab2015a}. The SD consists of 1660
autonomous water Cherenkov detectors (WCDs) arranged on a triangular grid with a spacing of
\unit[1500]{m}. Its geolocation is at $-69.0^\circ$ in longitude and $-35.4^\circ$ in latitude, in the
western part of Argentina. With a field of view to UHE photons limited to the zenith angle range between
$30^\circ$ and $60^\circ$ (as determined by data quality cuts necessary for the photon identification
method used), a fraction of 18.3\,\% of the whole sky is covered at any time. Due the field of view and 
the geolocation of the Observatory, 70.8\,\% of the sky is covered during a full rotation of the Earth. 
A small region with a radius of about $5^\circ$ around the celestial south pole is constantly observed.

The bulk of data received at the Pierre Auger Observatory originates from cosmic rays of hadronic
nature. With its different detector components and various enhancements, the Observatory is also
sensitive to a possible component of primary photons. Different searches have been performed
aiming to identify such a component among the diffuse flux of cosmic
rays~\citep{abreu22b,aab2017a,abreu22c}, as well as from steady point sources in the
sky~\citep{aab2017b}. No statistically significant excess of primary UHE photons has been identified
so far, and the strongest constraints to date on the flux of photons
from $2{\times}\unit[10^{17}]{eV}$ up to energies beyond
$\unit[10^{20}]{eV}$ have been obtained.

\begin{figure}[h]
 \center
 \includegraphics[width=\columnwidth]{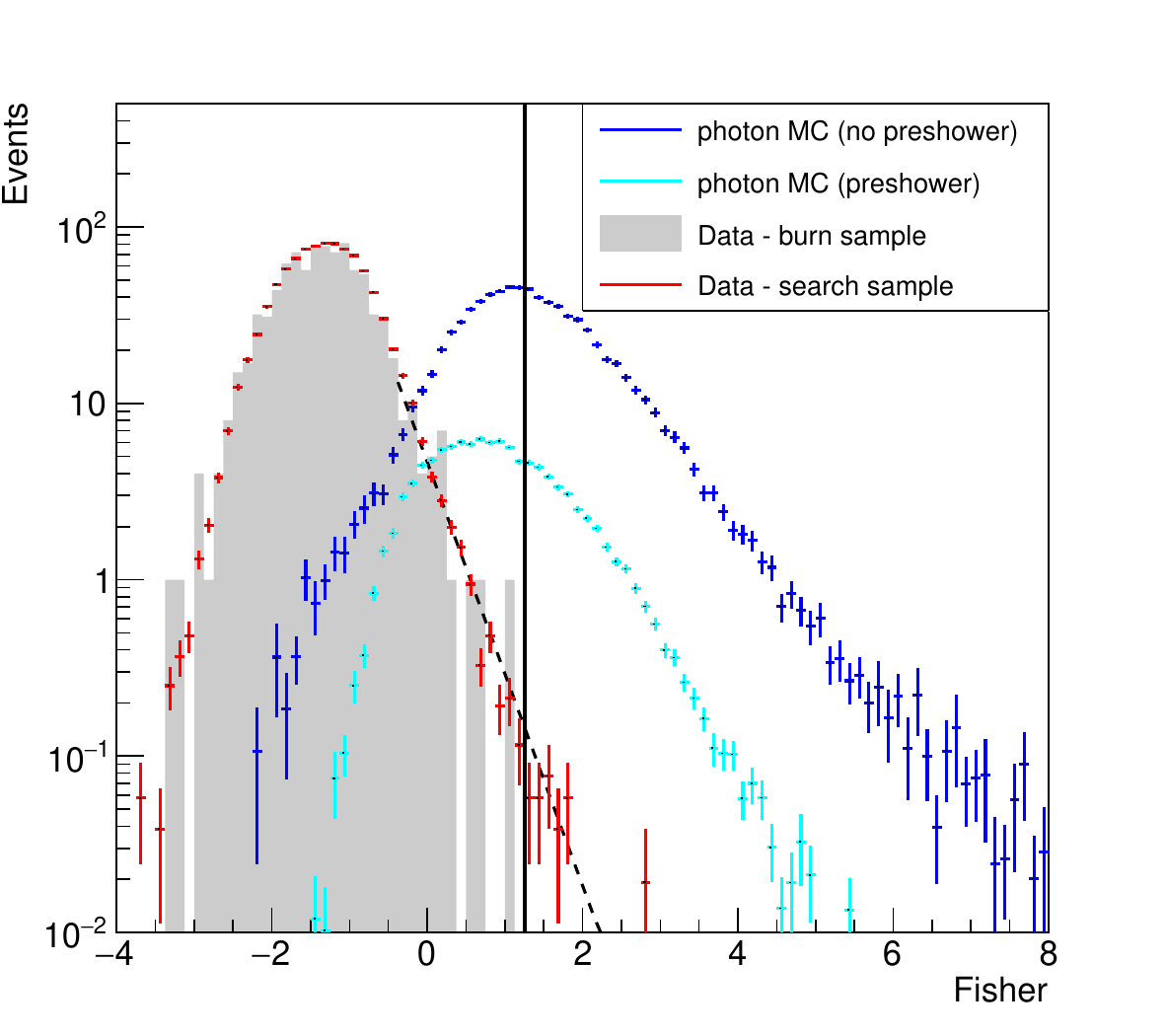}
 \caption
 {
   Distributions of the Fisher discriminant values of a set of simulated photon events (blue) and data
   events (red) with photon energies above $\unit[10^{19}]{eV}$ recorded by the
   SD~\citep{abreu22b}. The dark blue distribution shows the subset of photon simulations which did
   not initiate a preshower, i.e.\ did not interact in the geomagnetic field before reaching the atmosphere, 
   while the light blue distribution displays preshower events exclusively. This
   subset was used in~\cite{abreu22b} to derive the photon cut above which an event can be regarded 
   as a photon candidate (black vertical line). The right tail of the data distribution has been fitted by an 
   exponential function (tilted black line) to compare the number of observed events passing the cut 
   value with the expectation. The search sample and the photon distributions are scaled so as to have the same integral as the burn sample one (gray).
 }
 \label{fig:sd_analysis}
\end{figure}

If an air shower event in coincidence with a GW is found, a method is needed to judge the likelihood of it
originating from a primary photon or hadron. For this purpose, the photon discrimination method
from~\citet{abreu22b} is adopted which is briefly described in the following. This method utilizes the
data recorded by the SD taking advantage of its high duty cycle of almost 100\,\%. The identification of
photon-induced air showers is based on the shower lateral distribution, i.e. the distribution of particles
as a function of the distance to the shower axis, and the shapes of the signal time traces recorded by
the WCD stations. In particular, two discriminating photon observables are used, termed \lldf\, and 
$\Delta$. Photon-induced air showers, which are typically poor in muons, show on average a steeper
lateral distribution function (LDF) compared to hadron-induced showers. The observable \lldf\, measures
the signal in the WCDs as a function of their distance to the shower axis and is therefore sensitive to the
steepness of the LDF. The second observable, $\Delta$, quantifies the deviation of the risetime from a 
reference signal, typical of hadron-induced showers as measured in data.
It is sensitive to both the ratio between the electromagnetic and the muonic shower components at the
ground level, and to \xmax, which is the atmospheric depth (slant depth) where the shower reaches its maximum development.
Photon-induced air showers are expected to exhibit a large deviation from the average (hadronic) data, 
i.e large $\Delta$, because their signal risetime will be longer due to an intrinsically smaller muonic 
component and a less attenuated electromagnetic component (as a consequence of a deeper \xmax).
To maximize the
photon-hadron separation power, the observables are first normalized with respect to the total signal
and the direction of the shower axis and are then combined using a Fisher discriminant analysis. The
distributions of the Fisher discriminant of data events and a set of simulated photon events are shown in
Fig.~\ref{fig:sd_analysis}. A typical photon-induced air shower is expected to have a significantly larger
Fisher discriminant value than the average event found in data. The distributions shown in
Fig.~\ref{fig:sd_analysis} provide a measure with which to judge the likelihood of a single event 
originating from a primary photon.

On the axis of the Fisher discriminant, a threshold value may be placed to define which events will be
accepted as ``photon candidate events''. Depending on this photon candidate selection cut, the 
photon-discrimination method described above has a non-zero rate of expected false-positive detections
contributing a certain amount of background within the signal region. In~\cite{abreu22b}, out of all air
shower events recorded during a period of \unit[16.5]{yr} (2004 January $-$ 2020 June), 16 events
passed the photon candidate cut which was placed at the median of the distribution of photon
simulations in that analysis (c.f.\ vertical line in Fig.~\ref{fig:sd_analysis}). This number was found to
be consistent with the expected hadronic background.

The photon discrimination method is optimized for air showers with incident zenith angles $\theta$
between $30^\circ$ and $60^\circ$ and photon energies $E_\gamma > \unit[10^{19}]{eV}$. Since
the energy scale of the SD is calibrated using hadronic air showers observed by both the SD and the
fluorescence detectors, the energy of a possible photon-induced air shower would be underestimated. 
In order to obtain a less biased estimator for the photon energy $E_\gamma$, the hadronic energy 
scale has been replaced by a function of $S(1000)$ and $\theta$ calibrated with photon 
simulations~\citep{abreu22b}.  Here, $S(1000)$ is the interpolated average signal produced in an 
SD station with a perpendicular distance of \unit[1000]{m} from the shower axis. The photon energy 
estimator $E_\gamma$ can be calculated for any air shower with reconstructed $S(1000)$ and zenith 
angle and is used to define the lower energy cut for the application of the analysis.

To clean the shower data set of non-well reconstructed events, a number of selection criteria is imposed
prior to the calculation of the discriminating air shower observables. The selected events are required to
have a successfully reconstructed shower axis and LDF, and have to fulfill the 6T5 trigger criterion (= 6
active SD stations around the station with the highest signal). For the calculation of $\Delta$ and \lldf,
events with reconstructed hadronic energy $E_{\text{hd}} < \unit[10^{18}]{eV}$ (energy estimator
obtained by the standard SD energy reconstruction~\citep{aab2015a}), and events without triggered
stations (excluding stations with a saturated low-gain channel) more than \unit[1000]{m} away from
the shower axis are rejected. A more detailed description of the two observables, \lldf\, and $\Delta$,
and further details of the photon-hadron separation method can be found in~\cite{abreu22b}.

\section{GW data}\label{sec:gw_data}

The GW events considered in this analysis were recorded by Advanced LIGO and Virgo during their first
three observation runs and published in three gravitational wave transient catalogs: 
GWTC-1~\citep{gwtc1}, GWTC-2~\citep{gwtc2} with its second revision GWTC-2.1~\citep{gwtc2.1},
and GWTC-3~\citep{gwtc3}. While the first catalog covers the observations of the first two runs O1
(from 2015 September 12 to 2016 January 19) and O2 (from 2016 November 30 to 2017 August 25), the
third observation run has been split in two parts, O3a (from 2019 April 1 to 2019 October 1) and O3b
(from 2019 November 1 to 2020 March 27), with a maintenance break of 1 month in between. The
observations of each part of O3 have been released in separate catalogs, GWTC-2 (GWTC-2.1) and
GWTC-3 respectively.

Key information about GWs which is important for this analysis is the localization of their sources. This
information is distributed by the GW observatories in the form of probability density distributions
realized via pixelized skymaps (``localization maps'') in the HEALPix~\citep{gorski2005a}
segmentation scheme. The resolution of these maps varies between GW events and typically depends
on the overall localization quality of a source. In addition to the directional localization, a ``best fit''
estimator for the luminosity distance $D_\text{L}$ is also given for each source. In the case of
GW170817, additional information about the host galaxy, NGC 4993, is available. In this case, the
source is treated as a point source and the well constrained distance to the host galaxy is used instead
of the estimate provided by the GW measurement. The pronounced differences in localization qualities
and distances give rise to a priorization of sources in the context of this analysis.

\section{Gravitational Wave Event Selection}\label{sec:gw_event_selection}

Due to shower-to-shower fluctuations, photon-induced air showers cannot unambiguously be separated
from the bulk of showers with hadronic origin. Since the 16 photon candidate events found
in~\cite{abreu22b} are consistent with the expected hadronic background, this number may serve as
an estimate of the background rate for the present analysis. This leads to a directional-averaged
background rate of $\beta_\text{cand} = \unit[(1.86_{-0.45}^{+0.58}){\times}10^{-21}]{cm^{-2}
~s^{-1}~sr^{-1}}$. With this rate, the expected number of background events passing the photon candidate cut from all
91 GW sources for a 1 day search period within the 90~\% localization regions is $b = 0.017$. Hence, the
background hypothesis could only be rejected at a level of $\unit[1.67]{\sigma}$ (derived using the 
Feldman-Cousins method described in the following paragraph), should a coincident
shower with a Fisher discriminant above the photon median be detected. Hence, an actual photon event
could not be identified as such and had to be attributed to the hadronic background because of the high
background rate. Possible ways to reduce the total background for a set of GW events include a
reduction of the temporal and directional search windows. With an additional selection of GW events, the 
background contamination of the search regions can further be efficiently reduced, boosting the
sensitivity of the analysis.

The sensitivity to a possible signal of primary photons can be quantified through the confidence level (CL) at
which the background hypothesis can be rejected in the case of a detection. For a given number of
observed photon candidate events and a given background, two-sided confidence intervals for the true
expectation value can be obtained through the construction described by \cite{feldman98a} (FC).
Depending on the CL, the lower limit of this interval may or may not be equal to zero.
Thus, as a convenient measure for the sensitivity, we define $\photonsignificance$ as the lowest
confidence level at which the lower FC limit is still consistent with zero for the given background and a
measured number of one photon candidate event. Technically, this is done by calculating the FC confidence
interval for a given background $b$ (i.e., the expected number of coincident air shower events not associated with a transient event) and an assumed signal $s = 1$ (i.e., the number of actual photon events originating from a particular transient) and iterating through the CL until the lower
limit matches exactly 0, i.e. so that a slightly lower confidence level would lead to a lower limit $> 0$.
With this definition, $\photonsignificance$ only depends on $b$ and, naturally, a higher number of expected 
background events leads to a lower value of $\photonsignificance$ -- the confidence level at which the 
hypothesis of a photon candidate belonging to the hadronic background can be rejected.

In the following, the term ``photon candidate event'' shall be used for air shower events with a Fisher
discriminant value larger than the photon candidate cut value used in~\cite{abreu22b}. However, for
this analysis, this value is only relevant for the definition of the GW selection strategy described in the
following. The likelihood of an air shower event that coincides with a GW source originating from a 
primary photon from that source depends on multiple parameters, like the precise value of the Fisher
discriminant, the direction of the source, and its localization quality.

\begin{figure*}
 \center
 \includegraphics[width=\textwidth]{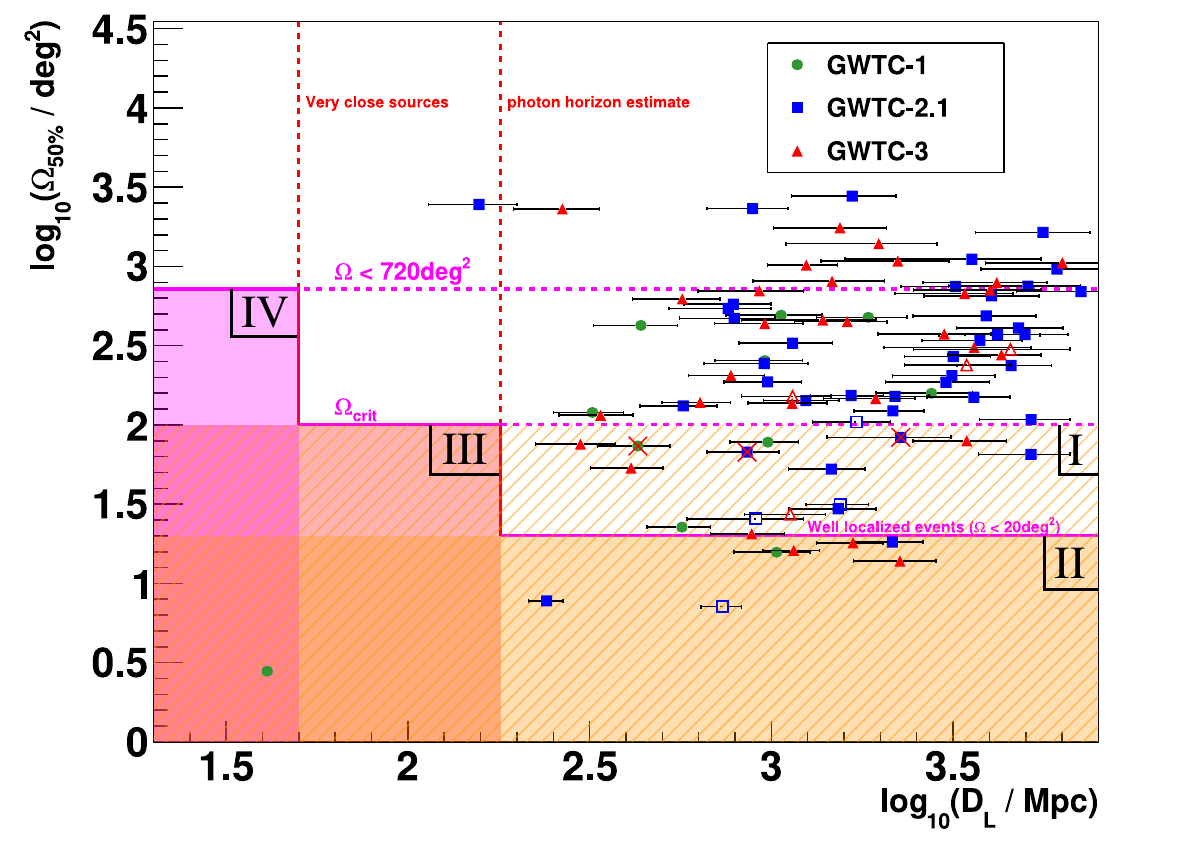}
 \caption
 {
   All GW events from GWTC-1 (green dots), GWTC-2.1 (blue squares) and GWTC-3 (red triangles) in the 
   space of source distance $D_L$ and localization $\Omega_{50\%}$. Events which are not within the
   field of view in the \unit[1]{day} time window are drawn with empty markers, while events which do
   at least partially overlap have solid markers. Three red crosses mark the events which pass the
   selection criteria for the short time window and also have an overlap with the field of fiew during that
   time. The shaded regions define the set of accepted events according to the selection citeria
   described in the text. The hatched region marks class~I which is solely relevant for the short analysis
   time window and the solid regions mark classes~II, III and IV.
 }
 \label{fig:omega_vs_dl}
\end{figure*}

As a first measure to limit the total background, two (mutually exclusive) time windows have been
defined during which a GW source is analzyed. A short time window of $\Delta t_{\text{short}} =
\unit[1000]{s}$ starting at $t_0 = \unit[-500]{s}$ before the GW event time, and a longer time window
$\Delta t_{\text{long}} = \unit[1]{day}$ starting at $t_0 = \unit[+500]{s}$ after the GW event time
have been chosen. While the short time window serves as a window for potential discovery with a high
degree of sensitivity, the long time window is the result of a compromise between sensitivity and a 
long-term follow-up and is loosely motivated by the time scale predicted by \cite{fang17a} for the
emission of UHE neutrinos.

An analysis of the GW sky localization maps distributed by LIGO/Virgo leads to the conclusion that using
their 50\,\% contour, defining a solid angle $\Omega_{50\%}$, as the search region in the sky is a
reasonable compromise between the expected level of background (which is proportional to the solid
angle of the analyzed sky region) and the confidence level at which the true source is localized within the
search region. By using the 50\,\% contour instead of the 90\,\% contour -- which is the most commonly
adopted convention -- on average about four times as many GW sources may be analyzed before the
same level of expected background is reached, while only losing 40\,\% in confidence that the source is
located within the analyzed sky region. To also take into account the directional resolution of the Auger
SD, which is about $1^\circ$ for photon-induced air showers above $\unit[10^{19}]{eV}$, the sky
localization maps of GW sources are convolved with a corresponding Gaussian distribution before 
constructing the 50\,\% contour.

In order to keep the sensitivity to a possible photon signal as high as possible, GW events are additionally
selected by their localization quality and distance. Close and well localized sources are preferred over
distant and poorly localized ones. Thus, optimal results can be obtained while keeping the expected
background at a reasonable level. Four classes of accepted GW events are defined here for which the 
50\,\% localization region is analyzed for coincident air shower events (see Fig.~\ref{fig:omega_vs_dl}).
These selection criteria can be summarized as

  $$ \begin{matrix}[l l l l ]
     (D_{\text{L}} < \infty & \text{and} & \Omega_{50\%} < \unit[100]{deg^2})_\text{s} & \text{``class~I''}\\
    & & & \\
     (D_{\text{L}} < \infty & \text{and} & \Omega_{50\%} < \unit[20]{deg^2})_\text{l} & \text{``class~II''}\\
    & & & \\
     (D_{\text{L}} < \unit[180]{Mpc} & \text{and} & \Omega_{50\%} < \unit[100]{deg^2})_\text{l} & \text{``class~III''}\\
    & & & \\
     (D_{\text{L}} < \unit[50]{Mpc} & \text{and} & \Omega_{50\%} < \unit[720]{deg^2})_\text{l,s} & \text{``class~IV''} \\ \smallskip
   \end{matrix} $$      
 with the lowercase ``l'' and ``s'' in the subscript in which time window (long and/or short) each
class of events is analyzed.
 
The first class (class~I) comprises GW sources with a maximum 50\% contour size of 
$\Omega_{50\%} = \Omega_\text{crit} = \unit[100]{deg^2}$ and any distance. The value of 
$\Omega_\text{crit}$ is
chosen such that $\photonsignificance$ for a photon candidate event within a \unit[1000]{s} time window
would always be above the $\unit[5]{\sigma}$ level in this specific event (i.e.,\ omitting any penalization
factor from multiple trials). Since classically no photon signal is expected from very distant sources, this
class also keeps a window open for potential discoveries of new physics. GW events in this class are
analyzed only in the short time window.

Especially well-localized sources with $\Omega_{50\%} \leq \unit[20]{deg^2}$ are additionally analyzed
in the long time window (class~II). From such a small region in the sky, the expected background would
still be small ($\photonsignificance > \unit[4]{\sigma}$) despite the longer observation time, and the
detection of a coincident photon-like event from a distant source could be a hint towards new physics.

\begin{deluxetable*}{lrrrrrrrr}
\tablehead{\colhead{} & \colhead{UTC time} & \colhead{$\delta_\text{GW}$} & \colhead{$\alpha_\text{GW}$} & \colhead{$D_L$ / $\unit[]{Mpc}$} & \colhead{$\Omega_{50\%}$ / $\unit[]{deg^2}$} & \colhead{source type} & \colhead{class} & \colhead{time window}}
 \tablecaption{A summary of the 10 GW events that pass the event selection described in the text. The 
 columns display (from left to right) the event identifier, the UTC time stamp of the GW detection by 
 LIGO/Virgo, the declination $\delta_\text{GW}$ and right ascension $\alpha_\text{GW}$ of the most
 likely source direction, the best estimate of the source luminosity distance $D_L$, the size 
 $\Omega_{50\%}$ of the 50\%-contour of the GW localization map after its convolution with the
 directional reconstruction uncertainty, and the most likely source type: binary black hole merger (BBH),
 binary neutron star merger (BNS) or black hole-neutron star merger (BHNS). The last two columns indicate the 
 classes in terms of $\Omega_{50\%}$ and $D_L$ that apply to each event and the time window of the present analysis during which 
 the 50\%-contour of the GW event was observed. For  further information on the individual GW data, the 
 reader may refer to the  corresponding catalogs~\citep{gwtc1,gwtc2,gwtc2.1,gwtc3} published by the
 LIGO/Virgo Collaborations and the
 associated public data release files (FITS files). \label{tab:gw_events}}
 \startdata
 GW150914               & 2015-09-14T09:50:45.4 & $-72.7^\circ$ & $-16.9^\circ$ & 429 & 73.6 & BBH & I & short \\
 GW170817               & 2017-08-17T12:41:04.4 & $ -23.4^\circ$ & $-162.6^\circ$ & 41 & 3.1 & BNS & all & long \\
 GW170818               & 2017-08-18T02:25:09.1 & $22.4^\circ$ & $-18.7^\circ$ & 1033 & 15.7 & BBH & I,II & long \\
 GW190517\_055101 & 2019-05-17T05:51:01.8 & $-46.5^\circ$ & $-130.9^\circ$ & 2270 & 83.6 & BBH & I & short \\
 GW190701\_203306 & 2019-07-01T20:33:06.6 & $-7.3^\circ$ & $37.8^\circ$ & 2152 & 18.2 & BBH & I,II & long  \\
 GW190728\_064510 & 2019-07-28T06:45:10.5 & $7.75^\circ$ & $-46.5^\circ$ & 858 & 67.3 & BBH & I & short \\
 GW190814               & 2019-08-14T21:10:39.0 & $-24.9^\circ$ & $12.7^\circ$ & 241 & 7.8 & BHNS & I,II & long \\
 GW200208\_130117 & 2020-02-08T13:01:17.9 & $-33.7^\circ$ & $139.4^\circ$ & 2258 & 13.8 & BBH & I,II & long\\
 GW200224\_222234 & 2020-02-24T22:22:34.4 & $-10.2^\circ$ & $175.2^\circ$ & 1677 & 18.0 & BBH & I,II & long\\
 GW200311\_115853 & 2020-03-11T11:58:53.4 & $-6.6^\circ$ & $1.6^\circ$ & 1152 & 16.2 & BBH & I,II & long \\
 \enddata
 \label{tab:gw_events}
\end{deluxetable*}

The long time window is also applied to GW events in the third class (class~III) which comprises 
sources with a maximum contour size of $\unit[100]{deg^2}$ which at the same time are required to 
be closer than \unit[180]{Mpc}. The maximum distance is chosen such that we reject GW sources 
from which no
photons are expected to reach the Earth even under the most optimistic assumptions about the photon
flux and its emission pattern, unless new physics is involved. For this choice, a ``photon horizon''
$h_\gamma$ has been estimated. This photon horizon is the distance up to which the energy transferred 
in to UHE by the so far brightest GW source, with a total radiated mass of almost $\unit[10]{M_\odot}$,
could be constrained to be less than its radiated GW energy. This distance is mainly driven by the photon
attenuation length in the extragalactic medium. Using the CRPropa 3 simulation
code~\citep{batista2016a} to simulate the propagation of UHE photons, a maximum horizon of
$h_\gamma = \unit[90]{Mpc}$ has been found for photons at $\unit[10^{20}]{eV}$. This horizon is
derived for isotropic emission. To take into account sources which might expose narrow jets pointing
directly towards Earth, only sources beyond $D_L > 2h_\gamma$ are rejected. 

A final class of accepted GW events (class~IV) allows especially close sources to be analyzed up to a
maximum allowed contour size of $\unit[720]{deg^2}$. For such sources with luminosity distance
$D_\text{L} \leq \unit[50]{Mpc}$, there is a realistic chance of observing a potential UHE photon flux or
at least placing strong physical constraints on the fraction of energy transferred into UHE photons. 
The value of $\unit[50]{Mpc}$ is defined by the maximum distance a source like GW170817 may 
have so that the fraction of energy transferred into UHE photons could still be constrained by a 
non-observation of photons at the SD array. The cut on the  maximum contour size is
chosen such that the bulk of GW events would be accepted and only the tail (about 10\%) in the
distribution of $\Omega_{50\%}$ is rejected, which can mostly be addressed to events which were not
observed by one of the two LIGO detectors. 

The four classes are not mutually exclusive and, hence, a single GW event may belong to multiple classes at the same time. Although classes II and III are subsets of class I in the space of $\Omega_{50\%}$ and $D_L$, an event belonging, e.g., to class II can only be analyzed in class I as well if its localization contour overlaps with the field of view during the short time window.

In Fig.~\ref{fig:omega_vs_dl}, the accepted regions in the space of source localization 
$\Omega_{50\%}$ and luminosity distance $D_\text{L}$ are visualized on top of the distribution of all 91
confident GW observations detected between O1 and O3b. In total, 23 GW events qualify in terms of $\Omega_{50\%}$ and $D_L$ for being checked in the short time window (classes I or IV), and a subset of 8 for also being checked in the long time window (classes II$-$IV).
Out of these 23 (8) GW events, in 3 (7) cases the
localization contours were at least partly covered by the
Auger SD field of view in the short (long) time window.
The 3 events in the short time window belong exclusively to class I, i.e. none of these events also qualifies for an inspection in the long time window.
All 7 events in the long time window are found in class II. One of these, GW170817, also passes the selection criteria for
classes I, III and IV, but it was not observable in the short time window. For a quick reference, further information about the 10 GW events that pass the event selection,
like the precise time stamp of their detection, the most likely source direction, the source distance and
most likely source type are compiled in Tab.~\ref{tab:gw_events}. For a more comprehensive reference
of the GW signals, one may refer to the official catalogs GWTC-1, GWTC-2, GWTC-2.1 and GWTC-3
published by the LIGO and Virgo Collaborations.

\section{Sensitivity}\label{sec:sensitivity}

In view of the future growth of the GW data set, let us first consider the overall sensitivity of this analysis
to a photon signal. As in Sec.~\ref{sec:gw_event_selection}, the sensitivity is quantified by adopting the
photon candidate cut value of \cite{abreu22b} and assuming a single photon candidate event within any
of the sky regions and time windows analyzed here for the 7(3) events selected. While the expected
number of random air showers (i.e.\ irrespective of the Fisher discriminant) to be coincident with any of
the analyzed GW sources and time windows is about 0.03, the expected total background with a Fisher
discriminant value exceeding the cut value  is $b = 9.1{\times}10^{-6}$ events. This leads to 
$\photonsignificance = \unit[4.44]{\sigma}$, meaning that the hypothesis of such a photon candidate
event belonging to the hadronic background could have been rejected at a CL of $\unit[4.44]{\sigma}$.
Since this value is calculated for the combined background from all selected GW sources, in both the long
and the short time windows, it naturally takes into account the trial factor that comes with an increasing
number of analyzed sources. Considering future applications of this analysis to larger sets of GW sources,
this penalized value of $\photonsignificance$ is expected to decrease. The real value of 
$\photonsignificance$ in the actual case of a coincident air shower detection will, however, strongly
depend on the precise values of the photon discrimination observables $L_\text{LDF}$ and $\Delta$ as
well as the direction and photon energy of the event. These values carry more detailed information about
the primary particle and its photon-likeliness than the binary selection method that is introduced by a
simple photon candidate cut.

\section{Results}\label{sec:results}

For both analysis time windows of \unit[1000]{s} and \unit[1]{day} the data of the Pierre Auger
Observatory have been analyzed for possible coincident photon events. No coincident air showers with
$E_\gamma > \unit[10^{19}]{eV}$ occured for any source in either of the time windows. This is well in
agreement with the expected amount of 0.03 chance coincidences. Consequently, also none of the 16
photon candidate events from~\cite{abreu22b} was found to be coincident with any of the selected
GWs. Following this non-observation of coincident events, for each GW source an upper limit on the
number of photons can be placed using the FC approach. In general, the FC upper limit at 90\% CL
without a measured signal and zero background is $N_\gamma^\text{UL} \approx 2.44$. The small
amount of background which is expected, however, does not significantly change this number.

From $N_\gamma^{\text{UL}}$, one can obtain limits on the corresponding
spectral photon flux $\frac{d\Phi_\gamma^{\text{GW}}}{dE_\gamma}(E_\gamma)$, which is the
number of photons arriving at the Earth from the direction of the GW source per unit time and area in the
energy range $[E_\gamma, E_\gamma+dE_\gamma)$. Assuming that the spectral photon flux follows a
power law with spectral index~$\alpha$, it can be written as
\begin{align}
\frac{\diff \Phi_\gamma^{\text{GW}}}{\diff E_\gamma} (E_\gamma) &= k_\gamma E_\gamma^{\alpha}. \label{eq:spectralflux}
\end{align}

With the energy dependence of the flux modeled as a power law, an upper limit on the flux normalization
factor $k_\gamma^{\text{UL}}$ can be derived from $N_\gamma^\text{UL}$ as 
\begin{align}
k_\gamma^{\text{UL}} &= \ufrac{N_\gamma^{\text{UL}}}{\int_{E_0}^{E_1} \diff E_\gamma E_\gamma^{\alpha}\mathcal{E} (E_\gamma,\theta_{\text{GW}}, \Delta t)}. \label{eq:fluxnorm}
\end{align}
For comparison to other results, the energy interval $[E_0,E_1)$ covers one order of magnitude starting
with $\unit[10^{19}]{eV}$. $\mathcal{E} (E_\gamma,\theta_{\text{GW}}, \Delta t)$ is the directional
exposure of the Observatory to photons with energy $E_\gamma$ within the time interval 
$\Delta t = t_1 - t_0$. The calculation of $\mathcal{E}$ is explained in the following.

\begin{figure}[h]
 \center
 \includegraphics[width=\columnwidth]{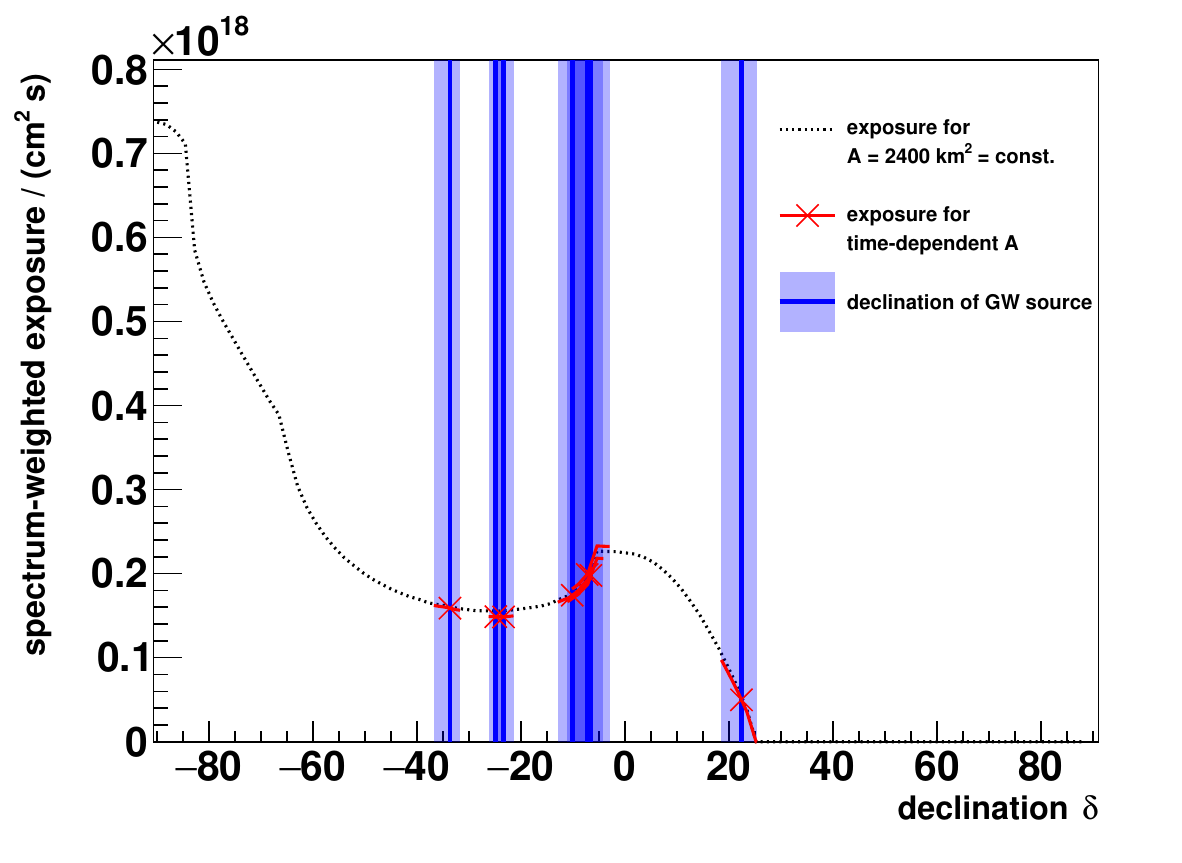}
 \caption
 {
   The spectrum-weighted exposure as a function of source declination for a benchmark  effective area of
   the SD array of $\unit[2400]{km^2}$ (dotted curve). The declination ranges covered by the 
   50\%-localization regions of the 7 GW events selected in the long time window are marked by the
   shaded bars with the most likely source directions marked by the dark blue bars. For each GW event the
   actual exposure, taking into account the time-dependent effective area of the SD in the long time
   window, is indicated by the red lines.
 }
 \label{fig:exposure}
\end{figure}

 The quality cuts that are imposed on the Auger SD data (see Sec.~\ref{sec:photon_at_auger}) limit the
 photon detection efficiency as a function of energy and zenith angle of the incident primary particle. The
 zenith-angle-averaged photon efficiency between $30^\circ$ and $60^\circ$ and 
 $E_\gamma > \unit[10^{19}]{eV}$, assuming an $E_\gamma^{-2}$ power-spectrum, has been found
 to be $\epsilon \simeq 0.54$. The efficiency has been derived using simulated photon events produced with 
 the CORSIKA simulation code~\citep{heck89a} and after applying the same selection cuts as used for data.
 With the photon efficiency given as a function of energy and direction, the exposure to UHE photons from 
 a transient point source at zenith angle $\theta_{\text{GW}}$ during the obervation period between $t_0 $ 
 and $t_1$ is given by
\begin{align}
 \mathcal{E}(E_\gamma,\theta_{\text{GW}},\Delta t) =& \int \limits_{t_0}^{t_1}~\diff t~A(t)~\epsilon(E_\gamma,\theta_{\text{GW}})\nonumber \\ &\times \Theta_{\text{FoV}}(\theta_{\text{GW}}(t))~\cos(\theta_{\text{GW}}(t)),\label{eq:exposure}
\end{align}
with $A(t)$ being the time-dependent effective area of the Auger SD array which is determined by the
number of active SD stations at a given moment. The step-function $\Theta_{\text{FoV}}$ accounts for
the fraction of the observation time in which the source is covered by the field of view of the SD between
zenith angles $30^\circ$ and $60^\circ$. Since the zenith angle $\theta_{\text{GW}}$ of a GW source is
a coordinate of the horizontal coordinate system which co-rotates with the Earth, $\theta_{\text{GW}}$
is a function of the sidereal time $t$, source right ascension $\alpha_{GW}$, and declination 
$\delta_{GW}$:
\begin{align}
 \cos(\theta_{\text{GW}}(t)) =& \sin\lambda \sin\delta_{\text{GW}}\nonumber \\ & + \cos\lambda \cos\delta_{\text{GW}} \sin(2\pi t/T - \alpha_{\text{GW}}),
\end{align}
with $\lambda$ being the latitude of the Auger SD array and $T$ the duration of a sidereal day. After
weighting the exposure by a $E^{-2}$-spectrum and integrating over a decade in energy, the 
spectrum-weighted exposure $\bar{\mathcal{E}}$ is a function of source right ascension and declination
during the short time window, while in the long time window $\bar{\mathcal{E}}$ depends only on
declination to first order and is depicted by the dotted curve in Fig.~\ref{fig:exposure}. The exposure has
a maximum at the celestial pole and vanishes for $\delta_\text{GW} > 24.6^\circ$. The basic structure of
the exposure curve is determined by the visibility of a certain direction in the zenith band between
$30^\circ$ and $60^\circ$ modulated by the directional photon detection efficiency $\epsilon$. For each
GW source analyzed in the long time window, the declination band covered by the 
$\Omega_{50\%}$-contour is highlighted in Fig.~\ref{fig:exposure} by a blue shaded bar with the most
likely source declination marked with a solid line. Since the effective area $A$ of the SD array varies over
time (typically only at the percent-level), the actual time-dependent values of the exposure are
highlighted in solid red next to the dotted benchmark-line, which is based on a fixed area of 
$\unit[2400]{km^2}$ corresponding to a typical average value.

\begin{figure*}
 \center
 \subfigure[]{\label{fig:fluence_upper_limits:1} \includegraphics[width=0.5\textwidth]{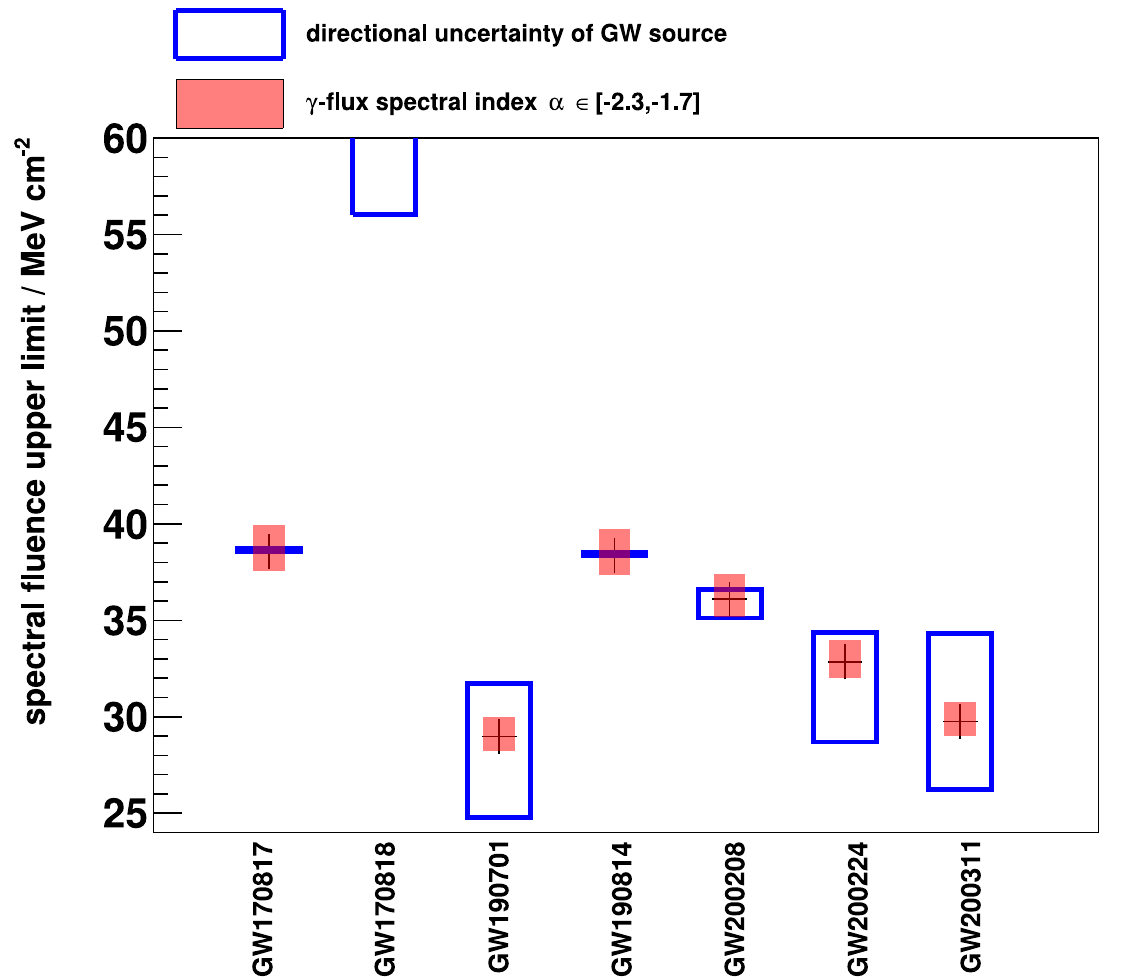}}%
 \subfigure[]{\label{fig:fluence_upper_limits:2} \includegraphics[width=0.5\textwidth]{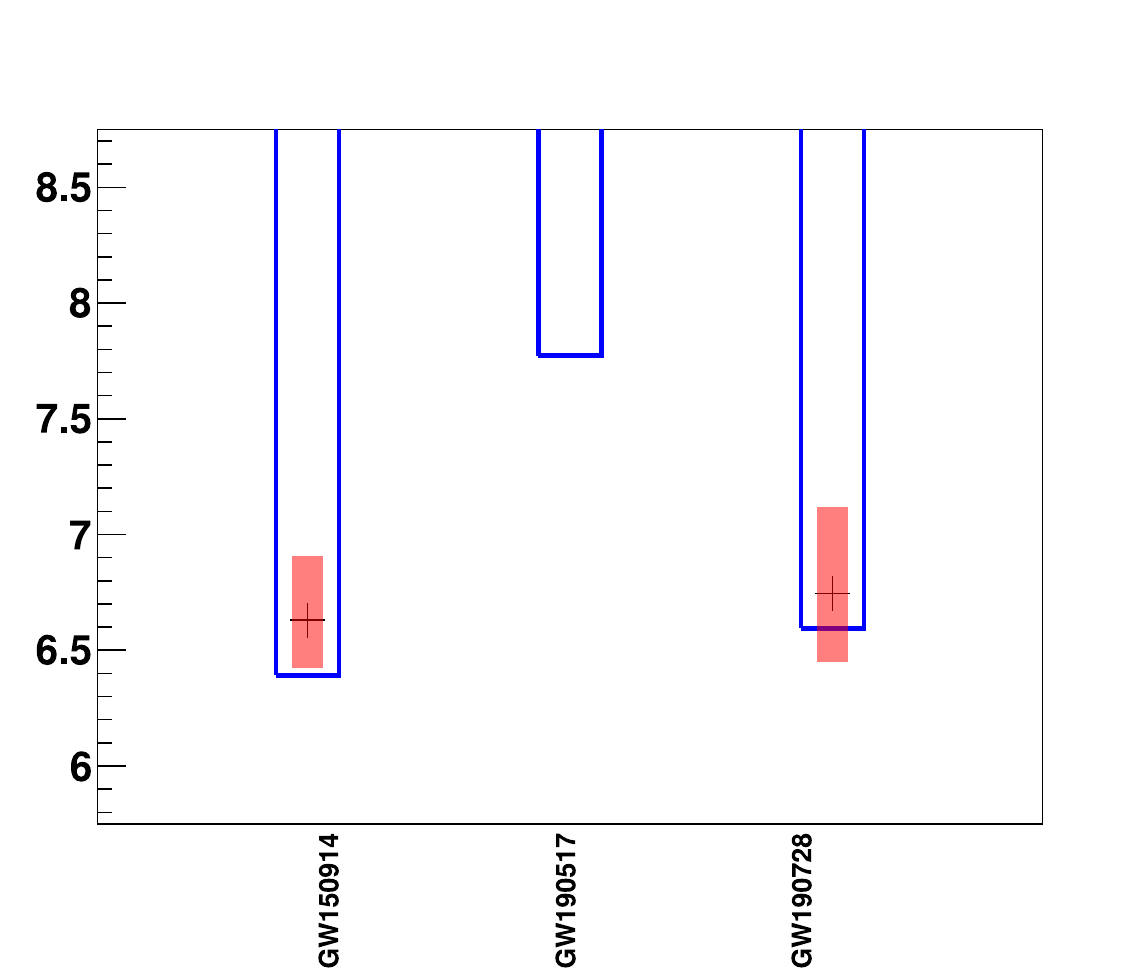}}
 \caption
 {
   Upper limits (at 90\% CL) on the spectral fluence of UHE photons from the selected GW sources for the
   searches in \subref{fig:fluence_upper_limits:1} the long and~\subref{fig:fluence_upper_limits:2} the
   short-time window. The limits for the most likely direction and a spectral index of $\alpha = -2$ are
   marked by the cross. The blue (empty boxes) error bars correspond to the variation of the upper limits
   due to the directional uncertainty of the source. Red (shaded boxes) error bars show the impact of a
   variation of the spectral index. For contours which are partly outside the field of view, the blue error
   bars grow to infinity (e.g.\ in the case of GW170818). While in the case of GW170818
   in~\subref{fig:fluence_upper_limits:1}, the most likely direction is close to the edge of the field of
   view, yielding a large upper limit of $\unit[109]{MeV~cm^{-2}}$, no limit could be placed on the most
   likely direction of GW190517 in~\subref{fig:fluence_upper_limits:2} as it was not inside the observed
   zenith angle range during the short time window.
 }
 \label{fig:fluence_upper_limits}
\end{figure*}

Finally, an upper limit on the spectral fluence $\fluence_\gamma^\text{UL}$ of UHE photons arriving
from a given source at the Earth can be derived from the flux upper limit:
 \begin{align}
     \fluence_\gamma^\text{UL} &= \int\limits_{t_0}^{t_1} \int\limits_{E_0}^{E_1} \diff t\,\diff E_\gamma\; E_\gamma\; \frac{d\Phi_\gamma^{\text{GW}}}{dE_\gamma}. \label{eq:specfluence1}
\end{align}
While no assumption on the time dependence of the flux is made, the extrapolation of the flux limits 
(which are based on data while the source is in the field of view) to the full time window implicitely 
assumes that the average flux during the period for which the source has been in the field of view 
is representative for the whole time window.  The limits on the spectral fluence depend on the exact 
direction of the GW 
source and change with a variation of the assumed spectral shape of the UHE photon flux. Hence, in
Fig.~\ref{fig:fluence_upper_limits} the results for $\fluence_\gamma^\text{UL}$ are shown for all
possible source directions within each localization contour and a variation of the spectral index 
$\alpha \in [-2.3, -1.7]$ for both time windows. In the long time window, all localization regions have been
fully covered by the field of view except for GW170818. Hence, this event could not be constrained for all
source directions within the $\Omega_{50\%}$-contour. All three GW sources in the short time window
have contours which partly leak out of the field of view. The upper limits that could be placed in the long
(short) time window vary typically around 
$\sim \unit[35]{MeV~cm^{-2}}$ ($\sim \unit[7]{MeV~cm^{-2}}$).

The BNS merger GW170817 plays a special role in this analysis for multiple reasons: as the first GW
source which has a confirmed observation of an electromagnetic counterpart, a
kilonova~\citep{arcavi2017a}, the BNS merger is especially interesting for all kinds of follow-up
multimessenger studies. So far, this source is also the only GW source for which the host galaxy has been
identified, in this case NGC\,4993 at a distance of about 
$\unit[41.0\pm 3.1]{Mpc}$~\citep{hjorth2017a}. This makes GW170817 the closest and best localized
source to date. While a large fraction of a potential UHE photon flux from NGC\,4993 is expected to be
attenuated by the cosmic background radiation fields, the intergalactic medium still has a degree of 
transparency to UHE photons and first constraints on the energy transferred into UHE photons 
can be derived. To accurately
take into account the interactions of photons, the photon attenuation has been studied as a function of 
photon energy using CRPropa 3. The upper limit to the spectral fluence at Earth 
$\fluence_\gamma^\text{UL}$, with an underlying flux modeled according to an $E^{-2}$-power-law spectrum, is 
then back-propagated to the source of the BNS and extrapolated to a full sphere to gain a limit on the energy
transferred into UHE photons. Upper limits that do not exceed the GW energy lower limit of
$E_\text{GW} \gtrsim \unit[0.04]{M_\odot}$ can be placed for
photons above  $\unit[2{\times}10^{19}]{eV}$. Furthermore, we find that the strongest limits can be placed for photon
energies above $\unit[4{\times}10^{19}]{eV}$, where less than 20~\% of the total GW energy at 90~\% CL is transferred into UHE photons. Since the attenuation 
of UHE photons follows an exponential law, this result indicates that the energy transferred into UHE photons 
by an even closer GW source, which might be observed in the near future, could likely be constrained well below 
the percent level.

\section{Conclusion}\label{sec:conclusion}

With the large exposure of its surface detector array, the Pierre Auger Observatory has been utilized to
investigate a possible outflow of ultra-high-energy photons from the gravitational wave sources detected
in recent years. The focus of this study were photons with energies above $\unit[10^{19}]{eV}$.
Searching for transient point sources of photons at such energies comes with two major difficulties: the
attenuation of ultra-high-energy photons in the cosmic background radiation fields which reduces the
photon interaction length to only a few Mpc, and the separation of primary photons from an overwhelming
background of hadronic cosmic rays using air shower properties. To overcome these obstacles, an
educated selection of gravitational wave sources has been defined aiming to maximize the physics impact
of the results. These -- in total 10 -- sources were analyzed for a coincident photon signal in a time span
ranging from \unit[500]{s} before the gravitational wave until one sidereal day after. Following the 
non-observation of a coincident signal, limits on the spectral fluence of photons in the respective energy
range were constructed assuming an $E_\gamma^{-2}$ power-law spectrum. These are the first limits on
UHE photons from GW sources.

The limits on the binary neutron star merger GW170817 add one further piece to the overall
multimessenger puzzle by constraining the electromagnetic outflow of the source in the UHE regime. The
results can be compared to the observed fluence of gamma rays between \unit[50]{keV} and 
\unit[300]{keV} as measured e.g.\ by the Fermi GBM~\citep{meegan2009a} to be 
$\unit[(2.8\pm 0.2){\times}10^{-7}]{erg~cm^{-2}}$~\citep{goldstein2017a} which are more than two
orders of magnitude stronger than the upper limits found here in the long time window (after
extrapolating the limits to a comparable range in $\log(E)$). The results can also be compared to the
limits on the photon flux between \unit[4]{TeV} and \unit[100]{TeV} placed by
HAWC~\citep{abeysekara2017a, abbott2017a} and on the fluence of neutrinos between 
\unit[100]{TeV} and \unit[1]{PeV} placed by IceCube~\citep{aartsen2017a, albert2017a}. After 
converting our limits to a comparable range in $\log_{10}(E)$, we find that
the limits placed in this work are of the same order of magnitude as the limits by HAWC (by a factor of  
$\sim 2.6$ weaker) and by a factor of $\sim 30$ stronger than the neutrino limits by IceCube. In the case
of HAWC, the comparable sensitivity of the observatories is mainly due to the exposure of the SD being
compensated by the higher expected particle flux at lower energies and HAWC's larger field of view, 
covering almost $\unit[2\pi]{sr}$. The difference to the IceCube
sensitivity additionally depends to a large extent on the different detection efficiencies between photons
and neutrinos.

With the upcoming GW observation run O4, starting prospectively in 2023, a further increase in the
detection rate is expected. With many more GW events to be analyzed in the future, a coincident air
shower from the cosmic ray background will be almost certain at some point. Then, the photon likeliness
of a coincident shower may be analyzed using dedicated simulations of photon-induced air showers
aiming to mimic the signal found in the data. Comparing the Fisher discriminant of a coincident shower
with the distributions obtained from photon simulations and hadronic background events, one
can then judge the overall photon likeliness of the air shower on an event-by-event basis.

This analysis is only a first step towards exploiting the full potential of the Pierre Auger Observatory in
multimessenger astronomy of transient point sources with UHE photons. While its sensitivity is already
competitive with that of other instruments measuring photons and neutrinos at lower energies,  the case
of GW170817 shows the potential of the Observatory if even closer GW sources should be detected in
upcoming observation runs. A future observation of e.g.\ a BNS merger in the Virgo cluster of galaxies
could possibly lead to a probe of the energy transferred into UHE photons at a level well below one percent of its GW
energy, a significant improvement compared to the $20\%~E_\text{GW}$
obtained in this work for GW170817.

\section*{Acknowledgments}

\begin{sloppypar}
The successful installation, commissioning, and operation of the Pierre
Auger Observatory would not have been possible without the strong
commitment and effort from the technical and administrative staff in
Malarg\"ue. We are very grateful to the following agencies and
organizations for financial support:
\end{sloppypar}

\begin{sloppypar}
Argentina -- Comisi\'on Nacional de Energ\'\i{}a At\'omica; Agencia Nacional de
Promoci\'on Cient\'\i{}fica y Tecnol\'ogica (ANPCyT); Consejo Nacional de
Investigaciones Cient\'\i{}ficas y T\'ecnicas (CONICET); Gobierno de la
Provincia de Mendoza; Municipalidad de Malarg\"ue; NDM Holdings and Valle
Las Le\~nas; in gratitude for their continuing cooperation over land
access; Australia -- the Australian Research Council; Belgium -- Fonds
de la Recherche Scientifique (FNRS); Research Foundation Flanders (FWO);
Brazil -- Conselho Nacional de Desenvolvimento Cient\'\i{}fico e Tecnol\'ogico
(CNPq); Financiadora de Estudos e Projetos (FINEP); Funda\c{c}\~ao de Amparo \`a
Pesquisa do Estado de Rio de Janeiro (FAPERJ); S\~ao Paulo Research
Foundation (FAPESP) Grants No.~2019/10151-2, No.~2010/07359-6 and
No.~1999/05404-3; Minist\'erio da Ci\^encia, Tecnologia, Inova\c{c}\~oes e
Comunica\c{c}\~oes (MCTIC); Czech Republic -- Grant No.~MSMT CR LTT18004,
LM2015038, LM2018102, CZ.02.1.01/0.0/0.0/16{\textunderscore}013/0001402,
CZ.02.1.01/0.0/0.0/18{\textunderscore}046/0016010 and
CZ.02.1.01/0.0/0.0/17{\textunderscore}049/0008422; France -- Centre de Calcul
IN2P3/CNRS; Centre National de la Recherche Scientifique (CNRS); Conseil
R\'egional Ile-de-France; D\'epartement Physique Nucl\'eaire et Corpusculaire
(PNC-IN2P3/CNRS); D\'epartement Sciences de l'Univers (SDU-INSU/CNRS);
Institut Lagrange de Paris (ILP) Grant No.~LABEX ANR-10-LABX-63 within
the Investissements d'Avenir Programme Grant No.~ANR-11-IDEX-0004-02;
Germany -- Bundesministerium f\"ur Bildung und Forschung (BMBF); Deutsche
Forschungsgemeinschaft (DFG); Finanzministerium Baden-W\"urttemberg;
Helmholtz Alliance for Astroparticle Physics (HAP);
Helmholtz-Gemeinschaft Deutscher Forschungszentren (HGF); Ministerium
f\"ur Kultur und Wissenschaft des Landes Nordrhein-Westfalen; Ministerium
f\"ur Wissenschaft, Forschung und Kunst des Landes Baden-W\"urttemberg;
Italy -- Istituto Nazionale di Fisica Nucleare (INFN); Istituto
Nazionale di Astrofisica (INAF); Ministero dell'Istruzione,
dell'Universit\'a e della Ricerca (MIUR); CETEMPS Center of Excellence;
Ministero degli Affari Esteri (MAE); M\'exico -- Consejo Nacional de
Ciencia y Tecnolog\'\i{}a (CONACYT) No.~167733; Universidad Nacional Aut\'onoma
de M\'exico (UNAM); PAPIIT DGAPA-UNAM; The Netherlands -- Ministry of
Education, Culture and Science; Netherlands Organisation for Scientific
Research (NWO); Dutch national e-infrastructure with the support of SURF
Cooperative; Poland -- Ministry of Education and Science, grant
No.~DIR/WK/2018/11; National Science Centre, Grants
No.~2016/22/M/ST9/00198, 2016/23/B/ST9/01635, and 2020/39/B/ST9/01398;
Portugal -- Portuguese national funds and FEDER funds within Programa
Operacional Factores de Competitividade through Funda\c{c}\~ao para a Ci\^encia
e a Tecnologia (COMPETE); Romania -- Ministry of Research, Innovation
and Digitization, CNCS/CCCDI UEFISCDI, grant no. PN19150201/16N/2019 and
PN1906010 within the National Nucleus Program, and projects number
TE128, PN-III-P1-1.1-TE-2021-0924/TE57/2022 and PED289, within PNCDI
III; Slovenia -- Slovenian Research Agency, grants P1-0031, P1-0385,
I0-0033, N1-0111; Spain -- Ministerio de Econom\'\i{}a, Industria y
Competitividad (FPA2017-85114-P and PID2019-104676GB-C32), Xunta de
Galicia (ED431C 2017/07), Junta de Andaluc\'\i{}a (SOMM17/6104/UGR,
P18-FR-4314) Feder Funds, RENATA Red Nacional Tem\'atica de
Astropart\'\i{}culas (FPA2015-68783-REDT) and Mar\'\i{}a de Maeztu Unit of
Excellence (MDM-2016-0692); USA -- Department of Energy, Contracts
No.~DE-AC02-07CH11359, No.~DE-FR02-04ER41300, No.~DE-FG02-99ER41107 and
No.~DE-SC0011689; National Science Foundation, Grant No.~0450696; The
Grainger Foundation; Marie Curie-IRSES/EPLANET; European Particle
Physics Latin American Network; and UNESCO.
\end{sloppypar}

\bibliography{bibliography}
\bibliographystyle{aasjournal}

\end{document}